# Spinful topological phases in acoustic crystals with projective *PT* symmetry

Yan Meng[1,#], Shuxin Lin[1,#], Bin-jie Shi[2,#], Bin Wei[3,4], Linyun Yang[1], Bei Yan[1], Zhenxiao Zhu[1], Xiang Xi[1], Yin Wang[2], Yong Ge[2], Shou-qi Yuan[2], Jingming Chen[1], Guigeng Liu[5], Hongxiang Sun[2,6,*], Hongsheng Chen[7], Yihao Yang[7,†], Zhen Gao[1,‡]

[1]*Department of Electrical and Electronic Engineering, Southern University of Science and Technology, Shenzhen 518055, China*
[2] *Research Center of Fluid Machinery Engineering and Technology, Faculty of Science, Jiangsu University, Zhenjiang 212013, China*
[3]*SKLSM, Institute of Semiconductors, Chinese Academy of Sciences, Beijing 100083, China.*
[4]*Center for Excellence in Topological Quantum Computation, University of Chinese Academy of Sciences, Beijing 100190, China*
[5]*Division of Physics and Applied Physics, School of Physical and Mathematical Sciences, Nanyang Technological University, 21 Nanyang Link, Singapore 637371, Singapore*
[6]*State Key Laboratory of Acoustics, Institute of Acoustics, Chinese Academy of Sciences, Beijing 100190, China*
[7] *Interdisciplinary Center for Quantum Information, State Key Laboratory of Modern Optical Instrumentation, ZJU-Hangzhou Global Science and Technology Innovation Center, College of Information Science and Electronic Engineering, ZJU-UIUC Institute, Zhejiang University, Hangzhou 310027, China.*
[#]*These authors contributed equally:* Yan Meng, Shuxin Lin, *and* Bin-jie Shi.

For the classification of topological phases of matter, an important consideration is whether a system is spinless or spinful, as these two classes have distinct symmetry algebra that gives rise to fundamentally different topological phases. However, only recently has it been realized theoretically that in the presence of gauge symmetry, the algebraic structure of symmetries can be projectively represented, which possibly enables the switch between spinless and spinful topological phases. Here, we report the first experimental demonstration of this idea by realizing spinful topological phases in "spinless" acoustic crystals with projective space-time inversion symmetry. In particular, we realize a DIII-class one-dimensional topologically gapped phase characterized by a 2ℤ winding number, which features Kramers degenerate bands and Kramers pair of topological boundary modes. Our work thus overcomes a fundamental constraint on topological phases by spin classes.

The concepts of topology and symmetry have revolutionized many branches of physics, ranging from condensed matter physics [1-4] to cold atoms [5], photonics [6-8], acoustics [9-19], and mechanics [20], as manifested by the classification of topological phases of matter. A fundamental dichotomy for topological classification is whether the studied systems are spinful or spinless. Remarkably, under internal or space group symmetries, these two spin classes can exhibit distinct topological phases. A representative example is the space-time inversion symmetry *PT*, where time-reversal symmetry (*T*) and space inversion symmetry (*P*) are combined together [21,22]. For the spinful class, it satisfies $(PT)^2 = -1$, enforcing Kramers doubly degenerate bands [Fig. 1(a)] in the entire Brillouin zone (BZ) that further give rise to many intriguing topological phases including one-dimensional (1D) topological superconductors or insulators in class DIII [23,24]; by contrast, for the spinless class, it satisfies $(PT)^2 = 1$, dictating real non-degenerate band structures [Fig. 1(b)] that further lead to its own unique topological phases, such as real Dirac semimetals [25]. Therefore, the spin class seems to impose fundamental constraints on the possible topological phases that a physical system can realize.

However, recent theoretical advances have remarkably revealed that [26], in the presence of gauge symmetry, the above fundamental limitation can be broken, i.e., it is possible to realize spinful (spinless) topological phases previously unique in spinless (spinful) systems. The underlying mechanism is that, with gauge symmetry, the crystalline symmetries of a system should be projectively represented, which further fundamentally modifies the algebraic structure of the symmetry group. Particularly, with the gauge symmetry, the space inversion symmetry *P* will be projectively represented as $\mathcal{P} = GP$ [26-29], where *G* is a gauge transformation. Properly choosing *G*, the projective *PT* symmetry (i.e., $\mathcal{PT}$) can satisfy $(\mathcal{PT})^2 = 1$ for spinful systems [Fig. 1(c)], or $(\mathcal{PT})^2 = -1$ for spinless systems [Fig. 1(d)]. This surprisingly suggests that the projective *PT* symmetry can exchange the fundamental symmetry algebra of spinless and spinful systems, enabling the switching between spinless and spinful topological phases.

Here, we experimentally demonstrate this idea by realizing spinful topological phases in spinless acoustic crystals with projective *PT* symmetry for the first time. Acoustic crystals have hitherto provided a versatile platform to study various topological phases under the framework of quantum-classical analogies [30-39]. Moreover, the positive and negative couplings [40-45] can be easily implemented in acoustic crystals, which are essential to constructing a $\mathbb{Z}_2$ gauge field that can exchange the symmetry algebra between spinless and spinful systems. Particularly, we experimentally realize in an acoustic crystal a DIII-class 1D topologically gapped phase characterized by a 2ℤ winding number, an acoustic analogue of 1D *T*-invariant *p*-wave topological superconductor [24,46,47]. Such a topological phase features Kramers double-degenerate bands and Kramers pairs of Majorana-like topological boundary modes [49-51], which were previously unique in spinful systems. Moreover, we observe an unconventional topological phase transition and four topological interface states at the domain wall between two nontrivial topological acoustic crystals with opposite nonzero $2\mathbb{Z}$ winding numbers ($\nu = \pm 2$).

To experimentally observe spinful topological phases in a spinless system, we design and fabricate a 1D acoustic crystal, as shown in Fig. 2(a). The experimental sample is fabricated with a standard three-dimensional (3D) stereolithography technique, and the printing material (photosensitive resin) can be considered as hard walls, which enclose a hollow region filled with air. A unit cell of the acoustic crystal is schematically shown in Fig. 2(b), which consists of 8 cylindrical resonators with a height of $h = 20$ mm and a radius of $r_0 = 10$ mm, and twenty coupling tubes (with radii of $r_1$, $r_2$, $r_y$, and $r_z$). Coupling tubes with positive and negative coupling coefficients are indicated by blue and red colors, respectively. Mechanism of the positive and negative coupling coefficients is schematically illustrated in Fig. 2(c). A cylindrical acoustic resonator supports a dipole-like resonance mode ($p_x$) with an eigenfrequency of $f_0 = 8.65$ kHz. When two resonators are connected by a coupling tube, they couple with each other and form two coupled dipole modes, leading to the splitting of eigenfrequency $f_0$. The splitting strength corresponds to the coupling coefficient ($\kappa$), which can be extracted from the eigenfrequency difference of two coupled dipole modes. Specifically, when the in-phase coupled dipole mode has a higher (lower) eigenfrequency than that of the out-of-phase one, the coupling coefficient is $\kappa > 0$ ($\kappa < 0$). According to this mechanism, the sign of coupling coefficients $\kappa$ can be switched by changing the relative connection positions of the coupling tubes (see details in the Supplementary Materials).

The fabricated sample in Fig. 2(a) follows a 1D spinless tight-binding model with positive and negative couplings, as shown in Fig. 2(d). Each unit cell (indicated by a grey cube) has 8 sites (indicated by white spheres). The positive (negative) couplings between nearest-neighbor sites are indicated by blue (red) bonds and labelled as $t$ with different subscripts, where $t_y$ and $t_z$ correspond to the coupling coefficients along $y$ and $z$ directions, respectively; $t_{d,1}$ ($t_{o,1}$) and $t_{d,2}$ ($t_{o,2}$) are the intra- and inter-cell couplings along the $x$-direction, respectively. The Hamiltonian of the tight-binding model shown in Fig. 2(d) can be written as:

$$\mathcal{H}(k) = t_y \Gamma_{100} + t_z \Gamma_{301} + \sum_{\{s=d,o\}} \begin{bmatrix} 0 & u_s(k) \\ u_s^*(k) & 0 \end{bmatrix} \otimes M_s, \quad (1)$$

where $\Gamma_{\mu\nu\lambda}$ is a 8 by 8 matrix, defined as $\Gamma_{\mu\nu\lambda} = \rho_\mu \otimes \tau_\nu \otimes \sigma_\lambda$. $\rho_\mu$, $\tau_\nu$, and $\sigma_\lambda$ are Pauli matrices and $\mu, \nu, \lambda = 0, 1, 2, 3$. $M_s$ is a diagonal matrix, when its subscript "$s$" is written as "$d$" or "$o$", we have $M_d = \text{diag}(1,0,0,1)$ or $M_o = \text{diag}(0,1,1,0)$. $u_d(k) = t_{d,1} + t_{d,2}e^{-ik}$ and $u_o(k) = t_{o,1} + t_{o,2}e^{-ik}$.

With the Hamiltonian in Eq. (1), we then investigate the projective $PT$ symmetry of the designed configuration with spinless character. In our case, $\hat{T}$ and $\hat{P}$ operators can be defined as $\hat{T} = \hat{\mathcal{K}}$ and $\hat{P} = \Gamma_{111}\hat{I}$, respectively. When consider these two operators together, the space-time inversion symmetry satisfies $(PT)^2 = 1$, and both operators preserve the spin (see details in the Supplementary Materials). Particularly, the negative and positive couplings in our system can construct a $\mathbb{Z}_2$ gauge field with $\mathbb{Z}_2 = \{\pm 1\}$. In Fig. 2(e), we present the front view ($x$-$z$ plane) of the tight-binding model. It can be seen that within each plaquette (indicated by a dashed rectangular box), there are one negative and three positive coupling bonds, and thus encloses a $\pi$ gauge flux. Gauge transformation ($G$) on the $\mathbb{Z}_2$ gauge field is defined as $G = \Gamma_{003}$ and satisfies [26,52]:

$$[G, T] = 0, \{G, P\} = 0, G^2 = 1. \quad (2)$$

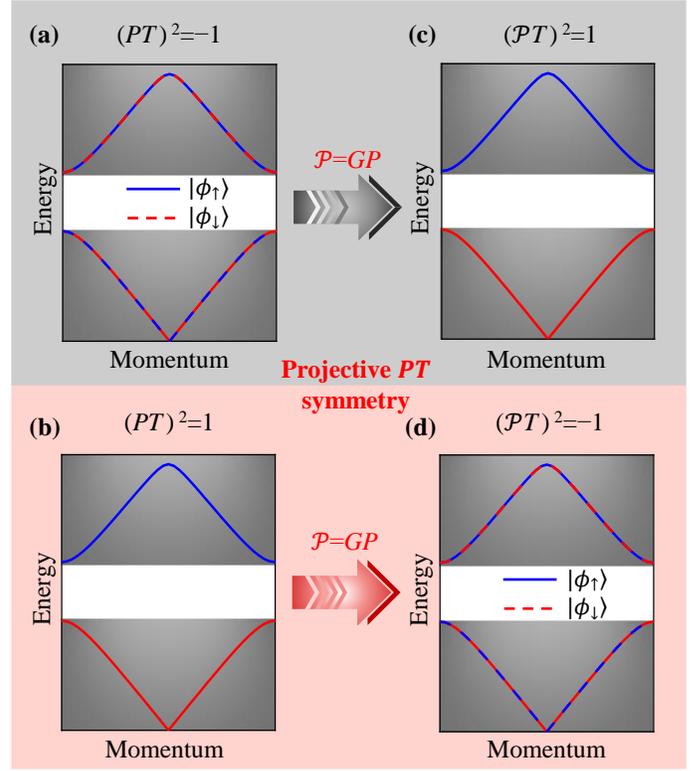

FIG. 1. Switching spinless and spinful topological phases with projective $PT$ symmetry. (a) $PT$ symmetry satifies $(PT)^2 = -1$ for spinful systems, enabling topological phases with Kramers double-degenerate band structures. (b) $PT$ symmetry satifies $(PT)^2 = 1$ for spinless systems, enabling topological phases with real non-degenerate band structures. (c), (d) Switching between spinful and spinless phases via projective $PT$ symmetry.

Here, the gauge transformation $G$ only swaps the negative and positive coupling coefficients along the $z$-direction, while keeping other coupling coefficients unchanged. The gauge transformation process is schematically illustrated in Fig. 2(e). It can be seen that the gauge flux in each plaquette is invariant under the gauge transformation $G$ (see details in the Supplementary Materials). In the presence of gauge symmetry, the space inversion symmetry operator $P$ can be projectively represented as $\mathcal{P} = GP$, thus $\mathcal{P}^2 = (GP) * (GP) = -(P)^2 = -1$. Hence, the projective $PT$ symmetry for the system satisfies $(\mathcal{P}T)^2 = -1$ (see details in the Supplementary Materials), indicating that the Kramers double-degenerate band structures previously unique to spinful systems are realizable in a spinless system under projective $PT$ symmetry.

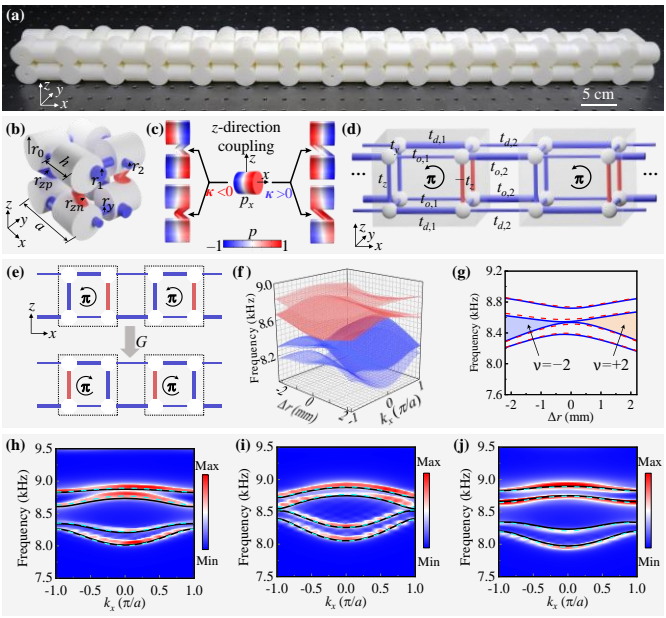

FIG. 2. Experimental demonstration of Kramers double-degenerate band structures. (a) Photograph of the fabricated 1D acoustic crystal sample with 15 unit cells along the $x$-direction. (b) Unit cell of the 1D acoustic crystal. Red (blue) tubes represent negative (positive) couplings. Lattice constant is $a$, height and radius of the cylindrical resonator are $h$ and $r_0$, radii of the coupling tubes along $x$, $y$, $z$ directions are $r_1$, $r_2$, $r_y$, $r_{zn}$, and $r_{zp}$, respectively. (c) Schematic illustration of positive ($\kappa > 0$) and negative ($\kappa < 0$) couplings along $z$-direction. (d) Configuration for the tight-binding model of the 1D acoustic crystal. Grey cubes: unit cell of the configuration. White spheres: sites. Blue (red) bonds: positive (negative) couplings. (e) Gauge transformation process (grey arrow) switches the sign of couplings in the $z$ direction, while keeping the synthetic gauge flux $\pi$ in each plaquette invariant. (f) Phase diagram of the simulated band structures of the 1D acoustic crystal as a function of $\Delta r$ and wave vector $k_x$. (g) Evolution of the middle bandgaps and topological winding numbers ($\nu = \pm 2$) as a function of $\Delta r$. (h)-(j) Measured (background colors) and simulated (black solid and cyan dashed lines) Kramers double-degenerate band structures of the 1D acoustic crystal with (h) $\Delta r = -2.2$ mm, (i) $\Delta r = 0$ mm and (j) $\Delta r = 2.2$ mm, respectively. The color scale indicates the measured acoustic energy density.

Next, we demonstrate the Kramers double-degenerate band structures of the designed 1D acoustic crystal. Without loss of generality, we set $r_y = 2.2$ mm, $r_{zn} = 2.7$ mm, $r_{zp} = 2.9$ mm, $(r_1 + r_2) = 6.2$ mm, and $(r_1 - r_2) = \Delta r$. To obtain the complete phase diagram, we numerically solve the band structures of the 1D acoustic crystals with a finite element method (commercial software COMSOL Multiphysics). We sweep $r_1$ from 2 mm to 4.2 mm (i.e., sweep $\Delta r$ from -2.2 to 2.2 mm), and plot the phase diagram as a function of $\Delta r$ and wave vector $k_x$ in Fig. 2(f). From the phase diagram, we can see that there exist four Kramers double-degenerate bands guaranteed by the projective $PT$ symmetry $(\mathcal{PT})^2 = -1$, and the targeted bandgap closes at $\Delta r = 0$ with a topological phase transition.

Hamiltonian in Eq. (1) can be decomposed into two off-diagonal blocks on account of its sublattice symmetry $S$. Therefore, the topological invariant, i.e., winding number ($\nu$) for the 1D acoustic crystal can be calculated by one of the two decomposed blocks [26]. As the designed 1D acoustic crystal satisfies $(\hat{\mathcal{P}}\hat{T})^2 = -1$ and $\{\hat{\mathcal{P}}\hat{T}, \hat{S}\} = 0$, where $\hat{S} = \Gamma_{333}\hat{I}$ is the sublattice symmetry operator, thus it is classified into D-III class topological phases with a winding number restricted to even integers $2\mathbb{Z}$ (see details in the Supplementary Materials), in contrast to the conventional Su-Schrieffer-Heeger (SSH) model characterized by a winding number of integers $\mathbb{Z}$. The evolution of frequencies as a function of $\Delta r$ at the BZ boundary (i.e., frequencies at $k_x = \pm\pi/a$) is shown in Fig. 2(g), which clearly shows the bandgap closure at $\Delta r = 0$. Winding numbers for systems with different $\Delta r$ are shaded by blue ($\nu = -2$) and orange ($\nu = +2$) colors, respectively. One can see that, as $\Delta r$ evolves from negative to positive, the winding number evolves from $\nu = -2$ to $\nu = +2$, and $\Delta r = 0$ is a critical point for the unconventional topological phase transition occurs between two nontrivial topological phases with opposite nonzero $2\mathbb{Z}$ winding numbers.

The evolvement of $\Delta r$ only changes the winding number from $\nu = -2$ to $\nu = +2$, while preserving the projective $PT$ symmetry of the proposed 1D acoustic crystals. Consequently, the Kramers double-degenerate band structures are guaranteed for all $\Delta r$. To experimentally demonstrate the Kramers double-degenerate band structures in acoustic crystals, we fabricate three samples with different $\Delta r$ (-2.2 mm, 0, and 2.2 mm, respectively) and apply spatial Fourier transform to the measured complex acoustic pressure distributions from real space to reciprocal space. As shown in Figs. 2(h-j), for each sample, four Kramers double-degenerate bands are observed in the entire BZ, and the measured results (background colors) are in good agreement with the simulated results (black solid and cyan dashed lines).

Then, we explore the Kramers pairs of topological boundary modes supported by the proposed 1D acoustic crystals. Based on the first-principles calculation, the proposed 1D acoustic crystals with $\Delta r \neq 0$ exhibit nontrivial spinful topological phases with nonzero $2\mathbb{Z}$ winding numbers. According to the bulk-edge correspondence, a nonzero winding number for spinful topological phases guarantees the existence of Kramers pairs of topological boundary modes. To verify the above prediction, we numerically calculate the eigenfrequency spectrum of a finite 1D acoustic crystal (15 unit cells) with $\Delta r = 2.2$ mm (see the detailed tight-binding model analysis in Supplementary Materials). As shown in Fig. 3(a), two pairs of topological boundary modes (colored dots) localized at the left or right ends can be observed within the bandgap of bulk states (grey squares), which are termed as Kramers pairs of topological boundary modes protected by spacetime inversion symmetry (see Supplementary Materials), in contrast to recently reported degenerate zero-energy topological states at disclinations, as a result of the preservation of chiral symmetry [53]. Inset in Fig. 3(a) shows the enlarged frequency regime of the four topological boundary modes. The simulated acoustic pressure distributions of the four topological boundary modes are plotted in Fig. 3(b), with each two of them localizing at the same end and forming a Kramers pair. Within each pair, two topological boundary modes exhibit distinct mode symmetries, as indicated by "+" and "−" signs at cross-sections CS1 and CS2 on the left boundary or CS3 and CS4 on the right boundary.

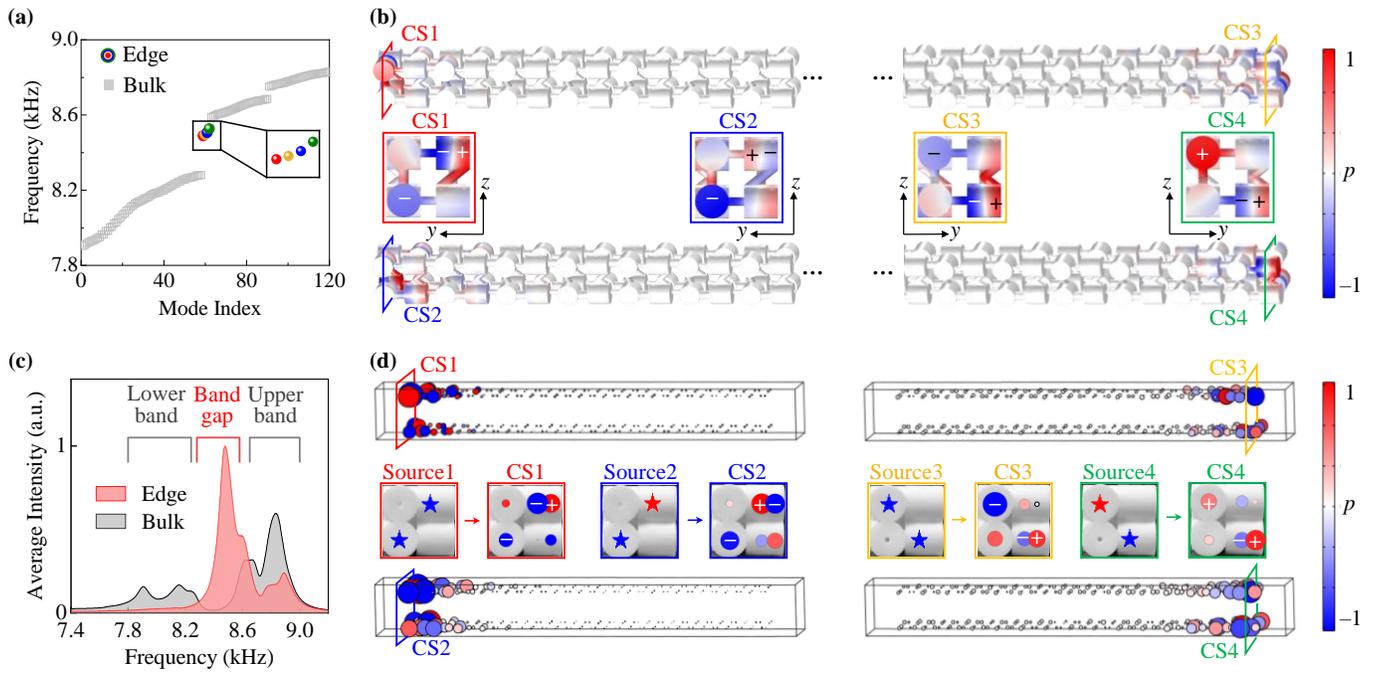

FIG. 3. Observation of Kramers pairs of topological boundary modes. (a) Simulated eigenfrequency spectrum of a finite-size 1D acoustic crystal (15 unit cells), eigenstates can be classified as bulk modes (grey squares) and boundary modes (color dots). Inset: enlarged frequency regime of boundary modes. (b) Simulated acoustic pressure distributions of two pairs of Kramers topological boundary modes. (c) Measured averaged intensity spectra for bulk (grey) and edge (red) regions, the simulated bandgap is highlighted by a red frame. (d) Measured acoustic pressure distributions of two pairs of Kramers topological boundary modes. Insets: Excitation sources (Source 1, Source 2, Source 3, and Source 4) and measured acoustic pressure distributions at four cross-sections (CS1, CS2, CS3, and CS4). Color stars indicate the position and phase of the excitation source. Color circles with "+" and "−" signs indicate the acoustic pressure distributions and symmetry of the excited boundary modes at each cross-section.

We then experimentally demonstrate the Kramers pairs of topological boundary modes. The sample is artificially divided into two non-overlapping parts: the "edge" region consists of 8 cylindrical resonators at both ends of the acoustic crystal, and the "bulk" region consists of the remaining 112 resonators. The average intensity is defined as the summation of normalized intensity for all resonators of the chosen region (either edge region or bulk region) averaged by the number of resonators in this region [42]. Specifically, acoustic intensity for a single resonator is measured by exciting the resonator in one hole and detecting the acoustic pressure in the other hole. The measured results are shown in Fig. 3(c), from which we can observe that within the bulk bandgap regime (about 8.3 to 8.5 kHz), the measured intensity spectrum of the edge region remains much higher than that of the bulk region, indicating the existence of edge states. Moreover, the measured peak frequency ranges of the edge and bulk regions match well with the simulated eigenfrequency spectrum shown in Fig. 3(a).

We also experimentally map the acoustic pressure distributions of the Kramers pairs of topological boundary modes. Using two acoustic point sources (indicated by red or blue stars) with designed phases difference (0 or $\pi$) placed at the left or right edges of the 1D acoustic crystal, we selectively excite each pair of topological boundary modes that possess different mode symmetries. Fig. 3(d) shows the measured acoustic pressure distributions of the Kramers pairs of topological boundary modes localized on the left or right edges of the 1D acoustic crystal, matching well with the simulated results shown in Fig. 3(b). Moreover, each pair of the measured topological edge states with their acoustic pressure localized on the same edge (left or right) exhibit almost the same eigenfrequency but different phase distributions (indicated by "+" and "−" signs), as shown in CS1 and CS2 (or CS3 and CS4). The experimental observations of Kramers pairs of topological boundary modes together with Kramers double-degenerate band structures, which are previously unique to spinful systems, demonstrate the spinful topological phases in spinless acoustic crystals. For comparison, the simulated average intensity and acoustic pressure distributions of topological edge states excited by a phased source array can be found in Supplementary Materials.

Finally, we demonstrate the topological interface states between two nontrivial spinful topological phases with opposite winding numbers ($\nu = \pm 2$). Fig. 4(a) shows the fabricated sample, which consists of 8 unit cells with winding number $\nu = -2$ on the left side and 8 unit cells with winding number $\nu = +2$ on the right side, and the red plane indicates the interface. According to the bulk-edge correspondence, the total number of topological interface states should be exactly equal to the difference between the winding numbers of two nontrivial topological acoustic crystals. Indeed, four topological interface states (colored stars) emerge within the bulk (grey squares) bandgap of the simulated eigenfrequency spectrum of the sample, as shown in Fig. 4(b). Using the pump and probe method, we measure the average intensity of the "interface" (red color) and "bulk" (grey color) regions, as shown in Fig. 4(c). We observe that the intensity peak of the interface region is approximately located within the bulk bandgap, implying the existence of topological interface states. Following a similar procedure used to probe the topological edge states, we plot the simulated (Fig. 4(d)) and measured (Fig. 4(e)) acoustic pressure distributions of the four topological interface states, which are localized in the vicinity of the sample interface and decay exponentially away from it.

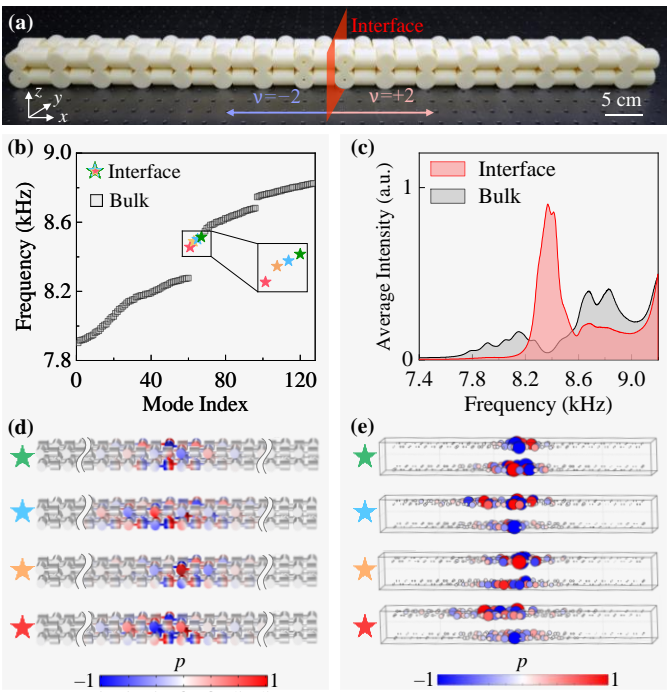

FIG. 4. Topological interface states between two spinful topological acoustic crystals with opposite topological invariants. (a) Photograph of the sample consists of two nontrivial acoustic crystals with opposite nonzero winding numbers $\nu = -2$ on the left side and $\nu = 2$ on the right side, respectively. Red plane represents the interface. (b) Simulated eigenfrequency spectrum for the configuration shown in (a). Inset: enlarged spectral region of four topological interface states (color stars). (c) Measured average intensity spectra for the "interface region" (red) and "bulk region" (grey), respectively. (d) Simulated and (e) measured acoustic pressure distributions of the four topological interface states.

We have thus experimentally realized spinful topological phases in "spinless" acoustic crystals for the first time, breaking the fundamental constraint on topological phase classification by the spin class. By introducing negative and positive couplings to engineer an equivalent $\mathbb{Z}_2$ gauge field in acoustic crystals, the projective $PT$ symmetry can completely modify the fundamental symmetry algebra of a system, making an originally spinless system behave like a spinful one. We observe the Kramers double-degenerate band structures and the Kramers pairs of topological boundary modes in acoustic crystals, which were previously unique to spinful systems. Though we choose acoustic crystals as a versatile platform for proof of principles, this protocol is general and can be directly extended to other spinless systems such as photonics, mechanics, and electronic circuits. This paradigm of projective symmetry reveals new perspectives on topological phase classification and opens the door to a vast landscape of unexplored topological physics with gauge degrees of freedom.

Z.G. acknowledge support from the National Natural Science Foundation of China under grant number 12104211, SUSTech Start-up Grant (Y01236148, Y01236248). The work at Zhejiang University was sponsored by the National Natural Science Foundation of China under grant number 62175215 and the Fundamental Research Funds for the Central Universities (2021FZZX001-19). H.S. acknowledged the support from the National Natural Science Foundation of China (Grant Nos. 11774137 and 12174159), and the State Key Laboratory of Acoustics, Chinese Academy of Science under Grant No. SKLA202016.

‡gaoz@sustech.edu.cn,
†yangyihao@zju.edu.cn
*jsdxshx@ujs.edu.cn